\documentstyle[12pt,a4]{article}

\newcommand{\limitstack}[2]{{\mbox{\scriptsize$\begin{array}[t]{@{}c@{}}%
~\\[-6.75ex]#1\\[-2.25ex]#2\end{array}$}}}
\begin{document}

\title{Solutions of Einstein's field equations related to Jacobi's
inversion problem}
\author{R. Meinel and G. Neugebauer 
\vspace{0.75cm}\\ Max--Planck--Gesellschaft\\
Arbeitsgruppe Gravitationstheorie  an der Universit\"at 
Jena\\ Max-Wien-Platz 1, D-07743 Jena, Germany\vspace {0.75cm}}
\date{}
\maketitle

\begin{abstract}
A new class of exact solutions to the axisymmetric 
and stationary vacuum Einstein equations containing $n$ arbitrary  
complex parameters and one arbitrary real solution of the axisymmetric 
three--dimensional Laplace equation is presented. The solutions are related to
Jacobi's inversion problem for hyper\-elliptic Abelian integrals. 
\end{abstract}

\newpage
\section{Introduction}
The inversion problem for hyperelliptic integrals (integrals over
a rational function $R[x,W(x)]$ with $W(x)$ being the square 
root of a polynomial
of order $2n+1$ or $2n+2$, $n = 2,3,4,\dots$) was formulated by Jacobi
in 1832 \cite{J}. Its solution in the ultraelliptic case ($n=2$)
was presented by G\"opel \cite{G} and
Rosenhain \cite{R}. They generalized 
Jacobi's theta functions to theta functions depending on two variables.
Riemann and Weierstrass solved the general problem in terms of theta functions
of $n$ variables, cf.~\cite{S}, \cite{K}, \cite{M}.

In the course of solving a physical problem within Einstein's theory of
general relativity \cite{NM1}, \cite{NM2} --- the problem of a 
rigidly rotating disk of dust ---
we were lead exactly to the ultraelliptic case of Jacobi's
inversion problem \cite{NM3}. In this letter we will show that a whole class
of solutions to the axisymmetric and stationary vacuum Einstein equations
may be associated with Jacobi's inversion problem in the general case.  
\section{The class of solutions}
As is well known the axisymmetric and stationary vacuum Einstein equations
may be reduced to the Ernst equation
\begin{equation}
(\Re f) \triangle f = (\nabla f)^2
\label{ernst}
\end{equation}
for the complex function $f(\rho,\zeta)$ called the 
Ernst potential \cite{E}, \cite{N}. The 
operators $\triangle$ and $\nabla$ have their usual three--dimensional
meaning applied to an axisymmetric function depending on the cylindrical
coordinates $\rho$ and $\zeta$:
\begin{equation}
\triangle = \frac{\partial^2}{\partial \rho^2} + \frac{1}{\rho}\frac{\partial}
{\partial \rho} + \frac{\partial^2}{\partial \zeta^2}, \quad
\nabla = (\frac{\partial}{\partial \rho},\frac{\partial}{\partial \zeta}).
\end{equation}

Any solution of our class is given by 
\begin{equation}
f = \exp\left\{\sum\limits_{m=1}^n \int\limits_{K_m}^{K^{(m)}}
\frac{K^n\,dK}{W} - u_n\right\} 
\label{class}
\end{equation}
with
\begin{equation}
W=\sqrt{(K+{\rm i}z)(K-{\rm i}\bar{z})\prod_{j=1}^n(K-K_j)(K-\bar{K}_j)}.
\label{W}
\end{equation}
The $K_m$  ($m=1,2,\dots,n$) are arbitrary complex constants; a bar denotes 
complex conjugation. The coordinates $\rho$ and $\zeta$ are combined to the
complex coordinate 
\begin{equation}
z=\rho+{\rm i} \zeta.
\end{equation}
The upper integration limits $K^{(m)}$ ($m=1,2,\dots,n$) are functions
of $\rho$ and $\zeta$ and have to be
determined as the solution of the following inversion problem: 
\begin{equation}
\sum\limits_{m=1}^n \int\limits_{K_m}^{K^{(m)}} \frac{K^j\,dK}{W} = u_j,
\quad j=0,1,2,\dots,n-1.
\label{inv}
\end{equation}
The quantities $u_j$ ($j=0,1,2,\dots,n$) in (\ref{inv}) and (\ref{class}) are
defined as follows:
Let $u_0$ be any real solution to the (axisymmetric) Laplace equation
\begin{equation}
\triangle u_0 = 0.
\end{equation}
Then  the $u_j$ ($j=1,2,\dots,n$) are real solutions to
\begin{equation}
{\rm i}u_{j,z} = \frac{u_{j-1}}{2} + z\,u_{j-1,z}, \quad j=1,2,\dots,n
\label{rel}
\end{equation} 
where an index ``$,z$'' denotes partial derivation with respect to $z$. As a
consequence, all $u_j$ are solutions to the Laplace equation. The calculation
of the $u_j$, starting from $u_0$, is simply a (path--independent) line
integration. If one starts from a solution $u_0$ of the Laplace equation that
vanishes at infinity one may choose the integration constants  
such that all $u_j$ vanish at infinity as well. In this case one obtains
$f\to 1$ at infinity, i.\,e.~an asymptotically flat solution to the Einstein
equations. It should be noted that the integrations in (\ref{class}) and
(\ref{inv}) have to be performed along the same curves in the two--sheeted
Riemann surface (of genus $n$) associated with $W(K)$ defined by (\ref{W}).
Using a result of Riemann [cf.~\cite{S}, p. 311, Eq.~(5)] it 
is possible to express $f$ in terms of theta functions of $n$ variables.
On the symmetry axis, $\rho = 0$, the two zeros $-{\rm i}z$ and ${\rm i}\bar{z}$
of the polynomial $W^2$ coincide. This leads, for $n\ge 2$, to 
expressions for the axis
values of the Ernst potential in terms of theta functions of $n-1$ variables.
Hence, the class of solutions considered cannot be obtained
by applying usual B\"acklund transformations repeatedly to an arbitrary 
Weyl metric. 

The proof that $f$ given by (\ref{class}) -- (\ref{rel}) satisfies indeed
the Ernst equation (\ref{ernst}) consists of two steps:

\noindent (i) Combining Eqs.~(\ref{rel}) and their complex conjugates with
(\ref{inv}) and (\ref{class}) one obtains
\begin{equation}
\sum\limits_{m=1}^n \frac{K^{(m)} + {\rm i}z}{W^{(m)}}\left(K^{(m)}\right)^j
K^{(m)}_{,z} = 0, \quad j=0,1,2,\dots,n-2,
\label{aa}
\end{equation}
\begin{equation}
\sum\limits_{m=1}^n \frac{K^{(m)} - {\rm i}\bar{z}}
{W^{(m)}}\left(K^{(m)}\right)^j
K^{(m)}_{,\bar{z}} = 0, \quad j=0,1,2,\dots,n-2,
\end{equation}
and
\begin{equation}
(\ln f)_{,z} = \sum\limits_{m=1}^n\frac{K^{(m)} + {\rm i}z}
{W^{(m)}}\left(K^{(m)}\right)^{n-1} K^{(m)}_{,z},
\end{equation}
\begin{equation}
(\ln f)_{,\bar{z}} = \sum\limits_{m=1}^n\frac{K^{(m)} - {\rm i}\bar{z}}
{W^{(m)}}\left(K^{(m)}\right)^{n-1} K^{(m)}_{,\bar{z}}
\label{zz}
\end{equation}
where 
\begin{equation}
W^{(m)}\equiv W\left(K^{(m)}\right).
\end{equation}

\noindent
Purely algebraic manipulations of Eqs.~(\ref{aa})--(\ref{zz}) lead to
\begin{equation} 
K^{(j)}_{,z} = (\ln f)_{,z}\frac{W^{(j)}}
{K^{(j)}+{\rm i}z}\prod_\limitstack{m=1}{(m\neq j)}^n 
\frac{1}{K^{(j)}-K^{(m)}},
\end{equation}
\begin{equation} 
K^{(j)}_{,\bar{z}} = (\ln f)_{,\bar{z}}\frac{W^{(j)}}
{K^{(j)}-{\rm i}\bar{z}}\prod_\limitstack{m=1}{(m\neq j)}^n \frac{1}
{K^{(j)}-K^{(m)}}.
\end{equation}

\noindent
The integrability conditions 
\begin{equation}
K^{(j)}_{,z\bar{z}}=K^{(j)}_{,\bar{z}z}
\end{equation}
yield, for all $j$,
\begin{equation}
\triangle (\ln f) = \alpha (\nabla \ln f)^2
\label{e2}
\end{equation}
with
\begin{equation}
\alpha = \sum\limits_{m=1}^n\left\{\frac{W^{(m)}}{(K^{(m)}+{\rm i}z)
(K^{(m)}-{\rm i}\bar{z})} \prod_\limitstack{j=1}{(j\neq m)}^n 
\frac{1}{K^{(m)}-K^{(j)}}\right\}.
\label{al}
\end{equation}

\bigskip

\noindent
(ii) Due to the reality of the $u_j$, it can easily be verified from 
Eqs.~(\ref{inv}) that the $K^{(m)}$ and $\bar{K}^{(m)}$ ($m=1,2,\dots,n)$
are related such that they form the $2n$ zeros of a characteristic
polynomial
\begin{eqnarray}
\prod_{j=1}^n (K-K_j)(K-\bar{K}_j) - (K+{\rm i}z)(K-{\rm i}\bar{z})
\left(\sum\limits_{j=0}^{n-1} \alpha_j K^j\right)^2 \nonumber \\
= (1-\alpha_{n-1}^2)\prod_{j=1}^n (K-K^{(j)})(K-\bar{K}^{(j)}).
\label{pol}
\end{eqnarray}
The (purely imaginary) quantities $\alpha_j$ may be determined from the
system of linear algebraic equations
\begin{equation}
\sum\limits_{j=0}^{n-1} \alpha_j(K^{(m)})^j = \frac{W^{(m)}}
{(K^{(m)}+{\rm i}z)(K^{(m)}-{\rm i}\bar{z})}, \quad m=1,2,\dots,n
\end{equation}
following from (\ref{pol}) for $K=K^{(m)}$ ($m=1,2,\dots,n)$ and
(\ref{W}). In particular, it turns out that
\begin{equation}
\alpha_{n-1} = \alpha
\end{equation}
with $\alpha$ defined according to (\ref{al}). Using (\ref{pol}) together
with (\ref{class}) one finds
\begin{equation}
\frac{f - \bar{f}}{f + \bar{f}} = \alpha_{n-1}
\end{equation}
and consequently, because of (\ref{e2}),
\begin{equation}
\triangle (\ln f) = \frac{f - \bar{f}}{f + \bar{f}}\, (\nabla \ln f)^2
\end{equation}
which is equivalent to the Ernst equation (\ref{ernst}).

\section{Discussion}
For $n=0$ the class of solutions (\ref{class}) reduces to Weyl's class with
$f=\exp(-u_0)$. The case $n=1$ leads to solutions in terms of elliptic
functions. A nontrivial and physically meaningful example for $n=2$ is
the solution of the problem of a rigidly rotating disk of dust \cite{NM3}.
This example provides some justification for the hope that the class
of solutions presented will lead to further physical applications in the
context of the rotating body problem.

An interesting mathematical question concerns the relation to Korotkin's
finite--gap solutions \cite{Ko}. It seems to us that those
solutions can be obtained for a particular choice of the potential
function $u_0$ and form therefore a subclass of the solutions
presented here.

\end{document}